**Asymptotic Behavior of Resonant Frequencies in Cylindrical Samples for Resonant Ultrasound Spectroscopy**

Jake E. Akins,[1] Casey M. Holycross[2], and Farhad Farzbod[1]

[1] *Department of Mechanical Engineering, University of Mississippi, University, MS 38677, USA*

[2] *Aerospace Systems Directorate (AFRL/RQTI), Wright-Patterson AFB, OH 45433, USA*

**ABSTRACT**: Resonant ultrasound spectroscopy (RUS) is a non-destructive technique for assessing the elastic and anelastic properties of materials by analyzing the frequencies of free vibrations in samples with known geometry. This paper explores the asymptotic behavior of eigenfrequencies in samples with cylindrical geometry. Extending prior research on cuboid samples, our study represents another step toward characterizing asymptotic behavior in arbitrarily shaped samples. While our findings are specific to cylindrical geometries, they are particularly relevant since many RUS samples adopt this shape. Furthermore, we present results on computing derivatives of Zernike polynomials, which may enhance the efficiency of resonant frequency calculations in the RUS method.

## I.     INTRODUCTION

Resonance Ultrasound Spectroscopy (RUS) is a non-destructive technique that uses the resonant vibrations of a sample to evaluate its material properties. Since its inception, RUS has been widely employed to characterize a diverse range of materials, including its main application in determining elastic constants [1-10], thin film properties [11, 12] crystallographic orientation [13, 14] and piezoelectric coefficients [15]. The versatility of RUS extends to its ability to measure elastic properties in a variety of materials, including bio materials such as human dentin [16], human bone [17], construction materials such as cement [18, 19], materials with initial strains and residual stress [20, 21], and elastomers [22]. Additionally, RUS can accommodate various geometries, boundary conditions, and temperatures, such as layered materials [23], cantilever boundary conditions [24] and complex geometries [25].

The use of resonance frequencies to detect defects dates back to British railroad engineers who tapped train wheels to listen for cracks [4]. The development of resonant ultrasound spectroscopy (RUS) offered a similar framework of nondestructive testing [4, 26]. With the increasing prevalence of additive manufacturing, RUS has become a valuable tool for assessing microstructural effects of heat treatment in 3D-printed metals [27] and for ensuring the quality of complex [28] and porous [29] additively manufactured components. Traditionally, RUS has utilized piezoelectric transducers for both excitation and detection of ultrasound [2, 5, 7, 30]. However, to minimize the coupling between transducers and the sample, non-contact methods such as laser-based techniques have been explored [8, 31-38]. These non-contact transducers enable effective RUS measurements without direct physical contact with the sample.

In RUS, material properties are determined by comparing experimentally measured resonant peaks with a theoretically calculated spectrum using an error function. The inputs to the spectral estimation are iteratively adjusted to minimize this error.

This paper concentrates on the theoretical aspects of resonant ultrasound spectroscopy (RUS), specifically investigating how rapidly the resonant frequencies increase. Understanding this theoretical growth is important because it sheds light on the amount of information present in higher frequencies. Beyond the inherent advantage of reducing experimental error by measuring more frequencies, it is uncertain whether additional experimental data actually provide more useful information. For example, certain off-diagonal elastic constants can have more pronounced effects at higher frequencies. In previous work [39], resonant frequencies for a cuboid sample approached an asymptote, indicating that beyond a certain point, increasing the number of experimental frequencies does not add new information. The ultimate aim is to find such asymptotic behaviors—if they exist—for samples of arbitrary geometry. In this paper, however, analysis is limited to cylindrical geometries. This study, together with earlier work on cuboid samples, serves as a stepping stone toward understanding the general case. While these findings are specific to cylindrical shapes, they are noteworthy due to the common use of cylindrical samples in typical RUS applications. Moreover, initial results on calculating the derivatives of Zernike polynomials are presented, which could potentially expedite calculations of resonant frequencies for cylindrical geometries.

This paper is structured as follows: First, the mathematical background essential for calculating resonant modes of free vibrations is presented. Next, Zernike polynomials are explored and the necessary results to compute the asymptotic behavior of resonant frequencies is shown. Finally, discussion of the insights obtained from studying this asymptotic behavior and examination of the influence of elastic constants.

## II. ASYMPTOTIC BEHAVIOR OF RESONANT FREQUENCIES

### A. RUS Calculation: Background

The methodology for calculating resonant frequencies in elastic solids is thoroughly detailed in the works of Demarest [1], Migliori et al. [5], and Visscher et al. [2]. These publications offer comprehensive explanations of a generalized approach that involves formulating the Lagrangian and identifying the frequencies and mode shapes that extremize it. This method effectively solves the elastic wave equation, enabling the computation of resonant frequencies in elastic materials. In this framework, kinetic and potential energy terms are detailed in Eq. (1), and the system's Lagrangian is presented in Equation (2). The displacement is assumed to be harmonic with frequency ω and is described using the mass density ϱ and displacement vector u. The Einstein summation convention is employed, where the indices i, j, k, l range from 1 to 3, representing the three spatial dimensions.

$$KE = \tfrac{1}{2}\rho\omega^2 u_i^2 \qquad PE = \tfrac{1}{2}C_{ijkl}\frac{\partial u_i}{\partial x_j}\cdot\frac{\partial u_k}{\partial x_l} \qquad (1)$$

$$L = \tfrac{1}{2}\int_V \left(\rho\omega^2 u_i^2 - C_{ijkl}\frac{\partial u_i}{\partial x_j}\cdot\frac{\partial u_k}{\partial x_l}\right)dV \qquad (2)$$

To approximate the displacements, we employ the Rayleigh-Ritz method with a finite set of basis functions. In this approach, the displacement in the *i* direction, denoted as $u_i$, is expressed in terms of selected basis functions $\varphi_q$:

$$u_i = a_{iq}\varphi_q \qquad (3)$$

Expanding and rearranging the Lagrangian leads to Equation (4). In this form, the volume integrals can be evaluated independently of the coefficients $a_{iq}$, allowing us to factor them out:

$$L = \tfrac{1}{2}a_{iq}a_{i'q'}\rho\omega^2\int_V \delta_{ii'}\varphi_q(x)\varphi_{q'}(x)dV - \tfrac{1}{2}a_{iq}a_{kq'}\int_V C_{ijkl}\frac{\partial\varphi_q}{\partial x_j}\cdot\frac{\partial\varphi_{q'}}{\partial x_l}dV \qquad (4)$$

The volume integrals involving the basis functions, their derivatives, and the elastic constants $C_{ijkl}$ can be computed independently of $a_{ij}$. We represent these integrals as matrices **E** and **Γ**, respectively. Thus, Equation (4) can be rewritten as:

$$L = \frac{1}{2}(\rho\omega^2 \boldsymbol{a}^T \mathbf{E}\boldsymbol{a} - \boldsymbol{a}^T \boldsymbol{\Gamma} \boldsymbol{a}) \tag{5}$$

where the coefficients $a_{iq}$ are assembled into vectors. Our objective is to find the values of $\boldsymbol{a}$ that extremize $L$. Using the Rayleigh-Ritz method, we set the derivative of $L$ with respect to $\boldsymbol{a}$ equal to zero:

$$\boldsymbol{\Gamma}\boldsymbol{a} = \rho\omega^2 \mathbf{E}\boldsymbol{a} \tag{6}$$

This equation represents a generalized eigenvalue problem. Although the polynomials suggested by Visscher et al. [2] are versatile and applicable to any geometry, our proof of the eigenfrequency's asymptotic behavior requires the matrix $\mathbf{E}$ in Eq. (6) to be the identity matrix. This condition means that the basis functions $\varphi_q$ must be orthonormal over the volume. Therefore, we utilize Zernike polynomials for the circular part of the cylindrical geometry [40], along with Legendre polynomials for the axial dimension of the cylindrical sample.

### B. Zernike Polynomials

In this study, we focus on cylindrical samples whose central axis aligns with the zzz-axis, with the origin of the coordinate system located at the midpoint of the cylinder's height. For these cases, Zernike polynomials are used for the disk coordinates, while Legendre polynomials are employed for the axis-aligned coordinate. Zernike polynomials, named after optical physicist and Nobel laureate Frits Zernike, are a set of orthogonal polynomials defined on the unit disk. They play a key role in diverse optics applications. Due to their orthogonality on the unit disk, it is appropriate to use Zernike polynomials here to set $\mathbf{E}$ as an identity matrix in Eq. (6). In polar coordinates, they are given by [41]:

$$\begin{cases} Z_\alpha^\beta(\rho, \theta) = R_\alpha^\beta(\rho) \cos(\beta\theta) \\ Z_\alpha^{-\beta}(\rho, \theta) = R_\alpha^\beta(\rho) \sin(\beta\theta) \end{cases} \tag{7}$$

Here, α and β are non-negative integers satisfying β ≤ α and α - β is even. To distinguish between sine and cosine terms, the negative sign is assigned to β in the definition of $Z$, not in $R$. In other words, the Zernike function is presented as $Z_\alpha^\mu(\rho, \theta)$ where μ=β for cosine terms and μ=-β for sine terms. Alternatively, the function can be defined with −α ≤ μ ≤α, β=|μ| and α- β is even. For the remainder of this paper, β is assumed to be a non-negative integer. Using μ, the Zernike functions include either sine or cosine terms. When referring specifically to the Zernike functions with cosine or sine terms, β and -β are used, respectively. This approach simplifies working with Zernike derivatives and related computations because the radial polynomials $R_\alpha^\beta(\rho)$ are the same for both sine and cosine terms. For simplicity and ease of notation, the radial polynomials $R_\alpha^\beta(\rho)$ (and consequently the Zernike functions) are zero for β > α. As an example, for α =4, the possible values for β are 0, 2, and 4, corresponding to five Zernike functions. For α=3, β takes the values 1 and 3, corresponding to four Zernike functions. The radial part of the Zernike polynomials is expressed as:

$$R_\alpha^\beta(\rho) = \sum_{S=0}^{\frac{(\alpha-\beta)}{2}} \frac{(-1)^S (\alpha-S)!}{S! \left[\frac{(\alpha+\beta)}{2}-S\right]! \left[\frac{(\alpha-\beta)}{2}-S\right]!} \rho^{\alpha-2S} \tag{8}$$

The orthogonality relations are given by:

$$\begin{cases} \int_0^1 R_\alpha^\beta(\rho) R_{\alpha'}^\beta(\rho) = \frac{\delta_{\alpha,\alpha'}}{2(\alpha+1)} \\ \int_0^{2\pi}\int_0^1 Z_\alpha^\mu(\rho,\theta) Z_{\alpha'}^{\mu'}(\rho,\theta)\, \rho d\rho d\theta = \frac{\epsilon_\beta \pi \delta_{\alpha,\alpha'} \delta_{\mu,\mu'}}{2(\alpha+1)} \end{cases} \tag{9}$$

with $\epsilon_\beta = 1 + \delta_{0,\beta}$. This means that $\epsilon_\beta = 2$ when $\beta = 0$ and $\epsilon_\beta = 1$ otherwise, accounting for the case where $\beta = 0$ and there are no sine or cosine terms. The cosine terms of the Zernike polynomials are plotted in Fig. (1) for various values of α and β, while the sine terms can be imagined as the cosine terms rotated by 90 degrees. It should be noted that Zernike polynomials, as presented in Eqs. (7) to (9), are not orthonormal, and are defined on a unit circle. These considerations are addressed later (section II-C) the expression for the basis functions.

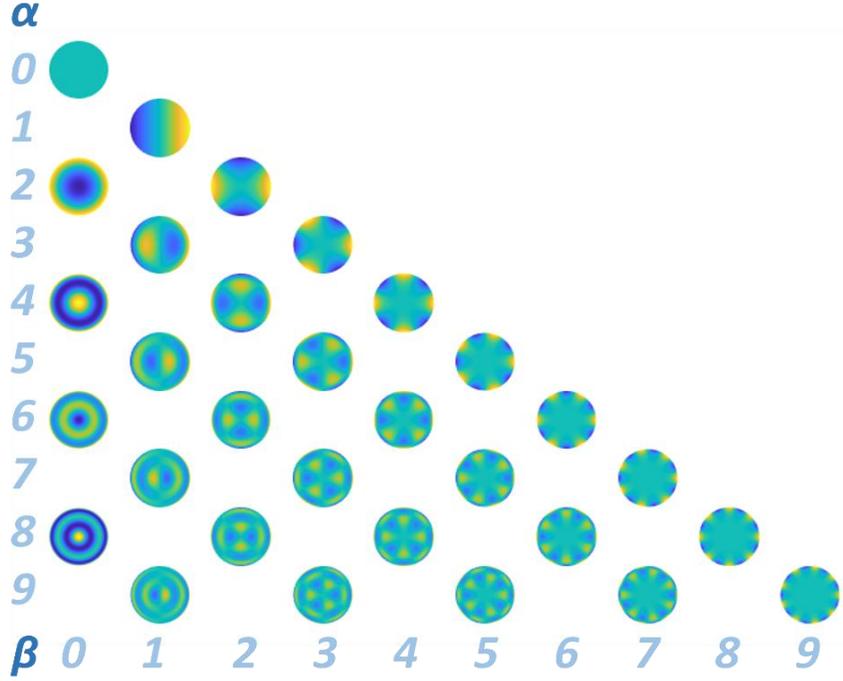

Figure 1: Zernike mode shapes for cosine terms and various α and β

### C. Asymptotic Analysis

The coordinate system is defined such that the base of the cylindrical sample lies in a plane parallel to the *xy*-plane with the central axis of the cylinder coinciding with the origin. The cylinder has a diameter D and a height $D_3$. For the basis function $\varphi_q$ in Eq. (3), two types of functions are used: Zernike's polynomials for the radial direction of the cylinder and Legendre polynomials for the axial direction. The basis function $\varphi_q$ as a function of the spatial variables $x_1$, $x_2$ and $x_3$ (along the *x*, *y*, and *z* axes, respectively) is expressed as:

$$\varphi_q(\mathbf{x}) = \left( \frac{Z_\alpha^\mu\left(\frac{2r}{D}, \theta\right)}{\sqrt{\epsilon_\beta \pi (8\alpha + 8)}} \right) \left( \frac{P_\gamma\left(\frac{2x_3}{D_3}\right)}{\sqrt{\frac{D_3}{2\gamma + 1}}} \right) , \quad \text{where} \quad \begin{cases} r = \sqrt[2]{(x_1)^2 + (x_2)^2} \\ \theta = \operatorname{atan2}(x_2, x_1) \\ -\frac{D_3}{2} < x_3 < \frac{D_3}{2} \end{cases} \quad (10)$$

where $P_\gamma$ is the Legendre polynomial of degree γ, and atan2($x_2$, $x_1$) is the two-argument arctangent function, which accounts for the correct quadrant of θ (i.e., angles greater than π/2). Alternatively, using complex notation, it can be expressed as arg($x_1 + ix_2$).

Similar to polynomial basis functions [2], a method to enumerate the basis functions is used. For the case of polynomial basis functions, $\varphi_q$ were defined as $x_1^k x_2^l x_3^m$, and the enumeration involved selecting N and considering all combinations of non-negative integers $(k,l,m)$ such that $k+l+m \leq N$. Here, however, the index the index μ is not independent of α (β=|μ| ≤ α and α-β is even). Therefore, the enumeration condition is defined as:

$$\alpha + \gamma \leq N \tag{11}$$

To visualize the range of (α, β', γ) that satisfy this inequality, consider N+3 squares arranged in a row, as shown in Fig. (2). Three of these squares are crossed out, effectively partitioning the arrangement into three segments. The number of uncrossed squares in each segment correspond to α, β', and γ, respectively.. Depending on the positions of the dashed lines, any of α, β', or γ can be zero. Since there are N remaining squares, α+γ ≤N.

A one-to-one relation between β' and β to define the sine and cosine terms of Eq. (7) is established. The relation is given by:

$$\mu = 2\beta' - \alpha \tag{12}$$

Here, the negative values of μ correspond to the sine terms in Eq. (7). For example, for two specific values of α, the mappings are:

$$\begin{cases} for\ \alpha = 4: \beta' = 0,1,2,3 \mapsto \mu = -4,-2,0,2,4 \\ for\ \alpha = 7: \beta' = 0,1,2,3,5,6 \mapsto \mu = -7,-5,-3,-1,1,\ 3,5,7 \end{cases} \tag{13}$$

This mapping shows how β values and sine and cosine terms are determined based on α and β'.

The total number of possible combinations of the basis function $\varphi_q$ in Eq. (10) can be determined, satisfying α+γ ≤N, by calculating the number of ways three dividing lines (crossing out squares) among N+3 squares. This is given by:

$$n = \frac{3(N+3)(N+2)(N+1)}{6} \tag{14}$$

In this formula, the denominator 6 (which is 3!) accounts for the indistinguishability of the dividing lines, and the 3 in the numerator is to account for $u_1$, $u_2$, and $u_3$. This number of combinations is similar to the case of polynomial basis functions. Because of the chosen orthonormal basis functions as defined in Eq. (10), the first integral in Eq. (4) vanishes when $q \neq q'$ and equals 1 when $q=q'$. Consequently, the matrix **E** becomes the identity matrix, and Eq. (6) simplifies to a standard eigenvalue problem:

$$\mathbf{\Gamma a} = \lambda \mathbf{a} \tag{15}$$

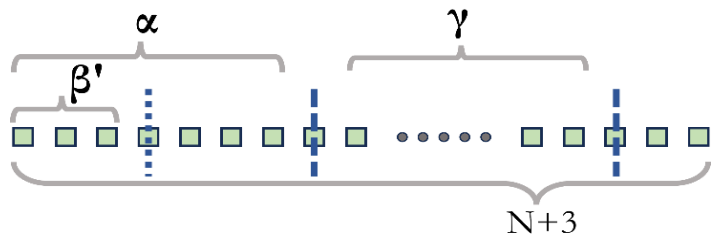

Figure 2: To determine all possible cases for $q$ under the condition α+β≤N, we consider N+3 squares. Among these, three squares are crossed out to divide the squares into three regions representing (α, β', γ). The value of α is the number of squares (excluding the crossed-out one) to the left of the middle-dashed line.

To further simplify the equations, $\varrho\omega^2$ is replaced with $\lambda$. Therefore, for the remainder of this paper, the eigenvalue $\lambda$ rather than the eigenfrequency $\omega$ is of primary focus.

The process of proving the existence and determining the value of the asymptotes of eigenfrequencies in cylindrical coordinates is similar to the approach used in the authors' previous work [39]. As demonstrated in [39], adding more basis elements leads eigenfrequencies with greater magnitude. The steps of the proof are as follows: [39]:

a) Analyze the trace of $\mathbf{\Gamma}$ in Eq. (6), which equals the sum of all its eigenvalues.

b) Finding the lower bound: Select N as specified in Eq. (11) yielding "n" basis functions according to Eq. (14). Consequently, the matrix $\mathbf{\Gamma}$ is an n×n square matrix with n eigenvalues. Note the largest of these eigenvalues, corresponding to the largest eigenfrequency, as $\lambda_n$.

Select N-1 to define a smaller set of basis functions, forming a smaller matrix $\mathbf{\Gamma}$, then compute the difference between the traces of the lower-dimensional $\mathbf{\Gamma}$ and the previous, larger-dimensional matrix. This difference corresponds to the average of the eigenvalues added as the matrix size increases. The $\lambda_n$ is bounded below by the average of these additional trace elements.

c) Finding the upper bound: Select N+1 to define an expanded basis set that results in a larger matrix $\mathbf{\Gamma}$. The average of the newly added eigenvalues corresponds to the difference between the traces of this new, higher-dimensional $\mathbf{\Gamma}$ and the previous one. Since the eigenvalues increase monotonically, this average exceeds $\lambda_n$.

d) From results in part (b) and (c), $\lambda_n$ is bounded between two values that converge to the same limit as N approaches infinity.

At this point, it is necessary to establish some properties of Zernike polynomials necessary for step **a** of the proof.

### 1. Some key properties of Zernike's polynomials

Zernike polynomials can be expressed in both polar and Euclidean coordinates. While integration is more conveniently performed in polar coordinates, the manipulation of elastic constants is simpler in Euclidean coordinates. Thus, the approach involves the following steps: first compute the necessary derivatives for calculating $\mathbf{\Gamma}$ (as defined in the second integral of Eq. (4)) in Euclidean coordinates, utilizing the following transformations between derivatives in polar and Euclidean coordinates:

$$\begin{cases} \frac{\partial}{\partial x} = \cos\theta \frac{\partial}{\partial \rho} - \frac{\sin\theta}{\rho} \frac{\partial}{\partial \theta} \\ \frac{\partial}{\partial y} = \sin\theta \frac{\partial}{\partial \rho} + \frac{\cos\theta}{\rho} \frac{\partial}{\partial \theta} \end{cases} \quad (16)$$

Next, employ the relations for derivatives of the radial components of the Zernike polynomials [41-43]. For β≥1, the following equations hold:

$$\begin{cases} \left(\frac{d}{d\rho} + \frac{\beta}{\rho}\right)\left(R_\alpha^\beta - R_{\alpha-2}^\beta\right) = 2\alpha R_{\alpha-1}^{\beta-1} \\ \left(\frac{d}{d\rho} - \frac{\beta}{\rho}\right)\left(R_\alpha^\beta - R_{\alpha-2}^\beta\right) = 2\alpha R_{\alpha-1}^{\beta+1} \end{cases} \quad (17)$$

These relations can be derived using the power series representation of the radial part of the Zernike function as given in Eq. (8). When β=0, the term β−1 (which equals to −1) in Eq. (17) is replaced by |β−1| (which equals to 1). The case of β≥1 is considered first and the case β=0 is addressed later. Adding and subtracting the equations in Eq. (17) yields [41]:

$$\begin{cases} \frac{d}{d\rho}\left(R_\alpha^\beta - R_{\alpha-2}^\beta\right) = \alpha\left(R_{\alpha-1}^{\beta+1} + R_{\alpha-1}^{\beta-1}\right) \\ \frac{\beta}{\rho}\left(R_\alpha^\beta - R_{\alpha-2}^\beta\right) = -\alpha\left(R_{\alpha-1}^{\beta+1} - R_{\alpha-1}^{\beta-1}\right) \end{cases} \quad (18)$$

Using Eq. (16), compute the partial derivatives of Zernike polynomials of Eq. (7), expressed in terms of cosine and sine components:

$$\frac{\partial z_\alpha^\beta}{\partial x} - \frac{\partial z_{\alpha-2}^\beta}{\partial x} = \cos\theta \left[\frac{d}{d\rho}\left(R_\alpha^\beta - R_{\alpha-2}^\beta\right)\right]\cos(\beta\theta) - \frac{\sin\theta}{\rho}\left(R_\alpha^\beta - R_{\alpha-2}^\beta\right)(-\beta)\sin(\beta\theta) \quad (19a)$$

$$\frac{\partial z_\alpha^{-\beta}}{\partial x} - \frac{\partial z_{\alpha-2}^{-\beta}}{\partial x} = \cos\theta \left[\frac{d}{d\rho}\left(R_\alpha^\beta - R_{\alpha-2}^\beta\right)\right]\sin(\beta\theta) - \frac{\sin\theta}{\rho}\left(R_\alpha^\beta - R_{\alpha-2}^\beta\right)(\beta)\cos(\beta\theta) \quad (19b)$$

$$\frac{\partial z_\alpha^\beta}{\partial y} - \frac{\partial z_{\alpha-2}^\beta}{\partial y} = \sin\theta \left[\frac{d}{d\rho}\left(R_\alpha^\beta - R_{\alpha-2}^\beta\right)\right]\cos(\beta\theta) + \frac{\cos\theta}{\rho}\left(R_\alpha^\beta - R_{\alpha-2}^\beta\right)(-\beta)\sin(\beta\theta) \quad (19c)$$

$$\frac{\partial z_\alpha^{-\beta}}{\partial y} - \frac{\partial z_{\alpha-2}^{-\beta}}{\partial y} = \sin\theta \left[\frac{d}{d\rho}\left(R_\alpha^\beta - R_{\alpha-2}^\beta\right)\right]\sin(\beta\theta) + \frac{\cos\theta}{\rho}\left(R_\alpha^\beta - R_{\alpha-2}^\beta\right)(\beta)\cos(\beta\theta) \quad (19d)$$

By employing Eq. (18), these can be further simplified into:

$$\frac{\partial z_\alpha^\beta}{\partial x} - \frac{\partial z_{\alpha-2}^\beta}{\partial x} = \cos\theta\cos(\beta\theta)\,\alpha\left(R_{\alpha-1}^{\beta+1} + R_{\alpha-1}^{\beta-1}\right) - \sin\theta\sin(\beta\theta)\,\alpha\left(R_{\alpha-1}^{\beta+1} - R_{\alpha-1}^{\beta-1}\right) \quad (20a)$$

$$\frac{\partial z_\alpha^{-\beta}}{\partial x} - \frac{\partial z_{\alpha-2}^{-\beta}}{\partial x} = \cos\theta\sin(\beta\theta)\,\alpha\left(R_{\alpha-1}^{\beta+1} + R_{\alpha-1}^{\beta-1}\right) + \sin\theta\cos(\beta\theta)\,\alpha\left(R_{\alpha-1}^{\beta+1} - R_{\alpha-1}^{\beta-1}\right) \quad (20b)$$

$$\frac{\partial z_\alpha^\beta}{\partial y} - \frac{\partial z_{\alpha-2}^\beta}{\partial y} = \sin\theta\cos(\beta\theta)\,\alpha\left(R_{\alpha-1}^{\beta+1} + R_{\alpha-1}^{\beta-1}\right) + \cos\theta\sin(\beta\theta)\,\alpha\left(R_{\alpha-1}^{\beta+1} - R_{\alpha-1}^{\beta-1}\right) \quad (20c)$$

$$\frac{\partial z_\alpha^{-\beta}}{\partial y} - \frac{\partial z_{\alpha-2}^{-\beta}}{\partial y} = \sin\theta\sin(\beta\theta)\,\alpha\left(R_{\alpha-1}^{\beta+1} + R_{\alpha-1}^{\beta-1}\right) - \cos\theta\cos(\beta\theta)\,\alpha\left(R_{\alpha-1}^{\beta+1} - R_{\alpha-1}^{\beta-1}\right) \quad (20d)$$

By applying trigonometric identities, the right-hand sides of Eq. (20) can be rewritten as follows:

$$\frac{\alpha}{2}[\cos(\beta+1)\theta + \cos(\beta-1)\theta]\left(R_{\alpha-1}^{\beta+1} + R_{\alpha-1}^{\beta-1}\right) - \frac{\alpha}{2}[\cos(\beta-1)\theta - \cos(\beta+1)\theta]\left(R_{\alpha-1}^{\beta+1} - R_{\alpha-1}^{\beta-1}\right) \quad (21a)$$

$$\frac{\alpha}{2}[\sin(\beta+1)\theta + \sin(\beta-1)\theta]\left(R_{\alpha-1}^{\beta+1} + R_{\alpha-1}^{\beta-1}\right) + \frac{\alpha}{2}[\sin(\beta+1)\theta - \sin(\beta-1)\theta]\left(R_{\alpha-1}^{\beta+1} - R_{\alpha-1}^{\beta-1}\right) \quad (21b)$$

$$\frac{\alpha}{2}[\sin(\beta+1)\theta - \sin(\beta-1)\theta]\left(R_{\alpha-1}^{\beta+1} + R_{\alpha-1}^{\beta-1}\right) + \frac{\alpha}{2}[\sin(\beta+1)\theta + \sin(\beta-1)\theta]\left(R_{\alpha-1}^{\beta+1} - R_{\alpha-1}^{\beta-1}\right) \quad (21c)$$

$$\frac{\alpha}{2}[\cos(\beta-1)\theta - \cos(\beta+1)\theta]\left(R_{\alpha-1}^{\beta+1} + R_{\alpha-1}^{\beta-1}\right) - \frac{\alpha}{2}[\cos(\beta+1)\theta + \cos(\beta-1)\theta]\left(R_{\alpha-1}^{\beta+1} - R_{\alpha-1}^{\beta-1}\right) \quad (21d)$$

These expressions can be further simplified to:

$$\alpha\left[R_{\alpha-1}^{\beta+1}\cos(\beta+1)\theta + R_{\alpha-1}^{\beta-1}\cos(\beta-1)\theta\right] \quad (22a)$$

$$\alpha\left[R_{\alpha-1}^{\beta+1}\sin(\beta+1)\theta + R_{\alpha-1}^{\beta-1}\sin(\beta-1)\theta\right] \quad (22b)$$

$$\alpha\left[R_{\alpha-1}^{\beta+1}\sin(\beta+1)\theta - R_{\alpha-1}^{\beta-1}\sin(\beta-1)\theta\right] \quad (22c)$$

$$-\alpha\left[R_{\alpha-1}^{\beta+1}\cos(\beta+1)\theta - R_{\alpha-1}^{\beta-1}\cos(\beta-1)\theta\right] \quad (22d)$$

These represent Zernike polynomials expressed in terms of cosine and sine components. Consequently, the partial derivatives of Zernike polynomials can be compactly represented as:

$$\frac{\partial z_\alpha^\beta}{\partial x} - \frac{\partial z_{\alpha-2}^\beta}{\partial x} = \alpha\left(z_{\alpha-1}^{\beta+1} + z_{\alpha-1}^{\beta-1}\right) \quad (23a)$$

$$\frac{\partial z_\alpha^{-\beta}}{\partial x} - \frac{\partial z_{\alpha-2}^{-\beta}}{\partial x} = \alpha\left(z_{\alpha-1}^{-(\beta+1)} + z_{\alpha-1}^{-(\beta-1)}\right) \quad (23b)$$

$$\frac{\partial z_\alpha^\beta}{\partial y} - \frac{\partial z_{\alpha-2}^\beta}{\partial y} = \alpha\left(z_{\alpha-1}^{-(\beta+1)} - z_{\alpha-1}^{-(\beta-1)}\right) \quad (23c)$$

$$\frac{\partial z_\alpha^{-\beta}}{\partial y} - \frac{\partial z_{\alpha-2}^{-\beta}}{\partial y} = -\alpha\left(z_{\alpha-1}^{\beta+1} - z_{\alpha-1}^{\beta-1}\right) \quad (23d)$$

Equations (23a-b) indicate that for the partial derivatives with respect to $x$, the cosine and sine terms remain consistent on both sides of the equations. In contrast, Equations (23c-d) demonstrate that for partial derivatives with respect to $y$, the roles of sine and cosine terms are interchanged. Variants of these equations, expressed in terms of complex numbers, have been previously explored by other researchers [44, 45]. However, as far as the authors are aware, the relationships derived in this paper are unique, specifically tailored for calculating the eigenfrequencies of RUS and are essential for the proof presented.

In the following section, expressions for the partial derivatives of Zernike polynomials are derived in terms of other Zernike polynomials. This process begins by examining Eq. (23a) and expanding it for progressively smaller values of $\alpha$:

$$\begin{cases}
\frac{\partial z_\alpha^\beta}{\partial x} - \frac{\partial z_{\alpha-2}^\beta}{\partial x} = \alpha\left(z_{\alpha-1}^{\beta+1} + z_{\alpha-1}^{\beta-1}\right) \\
\frac{\partial z_{\alpha-2}^\beta}{\partial x} - \frac{\partial z_{\alpha-4}^\beta}{\partial x} = (\alpha-2)\left(z_{\alpha-3}^{\beta+1} + z_{\alpha-3}^{\beta-1}\right) \\
\frac{\partial z_{\alpha-4}^\beta}{\partial x} - \frac{\partial z_{\alpha-6}^\beta}{\partial x} = (\alpha-4)\left(z_{\alpha-5}^{\beta+1} + z_{\alpha-5}^{\beta-1}\right) \\
\vdots \\
\frac{\partial z_{\alpha-2m}^\beta}{\partial x} - \frac{\partial z_{\alpha-2m-2}^\beta}{\partial x} = (\alpha-2m)\left(z_{\alpha-2m-1}^{\beta+1} + z_{\alpha-2m-1}^{\beta-1}\right) \\
\vdots
\end{cases} \quad (24)$$

In these equations, when $\alpha-2m=\beta$, the first term on the left-hand side remains nonzero, while the second term vanishes because $\alpha-2m-2<\beta$. Similarly, for this specific line, the first term on the right-hand side is also zero. Consequently, when summing over both sides of Eq. (24), all terms on the left-hand side cancel out except the first term. This leads to the following result:

$$\frac{\partial z_\alpha^\beta}{\partial x} = \left[\sum_{m=0}^{\frac{\alpha-\beta-2}{2}}(\alpha-2m)\left(z_{\alpha-2m-1}^{\beta+1} + z_{\alpha-2m-1}^{\beta-1}\right)\right] + \beta z_{\beta-1}^{\beta-1} \quad (25)$$

where the last term arises from the final nonzero equation in Eq. (24), corresponding to $2m=\alpha-\beta$. This equation demonstrates that the partial derivative of the Zernike polynomial with respect to $x$, for a given $\alpha$, can be expressed as a linear combination of Zernike polynomials with the same sine or cosine term but with smaller subscripts. This property is utilized later in section II-C-3 for calculating the integrals of partial derivatives over a given volume.

To proceed, the area integral of the squared partial derivative over the unit circle is computed, representing the norm of this derivative. Although the derivative is with respect to $x$, the area integral is performed in polar coordinates to simplify calculations using previously established relationships. The integral is expressed as:

$$\int_0^{2\pi}\int_0^1 \left(\frac{\partial z_\alpha^\beta}{\partial x}\right)^2 \rho d\rho d\theta = \int_0^{2\pi}\int_0^1 \left(\left[\sum_{m=0}^{\frac{\alpha-\beta-2}{2}}(\alpha-2m)\left(z_{\alpha-2m-1}^{\beta+1}+z_{\alpha-2m-1}^{\beta-1}\right)\right]+\beta z_{\beta-1}^{\beta-1}\right)^2 \rho d\rho d\theta \quad (26)$$

The right-hand side contains both squared terms and cross-multiplication terms. However, due to the orthogonality of Zernike polynomials, the cross terms vanish. Applying the norm definitions of Zernike polynomials over the unit circle, Eq 26 becomes:

$$\iint \left(\frac{\partial z_\alpha^\beta}{\partial x}\right)^2 \rho d\rho d\theta = \sum_{m=0}^{\frac{\alpha-\beta-2}{2}}\left[(\alpha-2m)^2\left\|z_{\alpha-2m-1}^{\beta+1}\right\|^2+(\alpha-2m)^2\left\|z_{\alpha-2m-1}^{\beta-1}\right\|^2\right]+(\beta)^2\left\|z_{\beta-1}^{\beta-1}\right\|^2 \quad (27)$$

Using Eq. (9) for the norm of Zernike polynomials yields:

$$\left\|z_{\alpha-2m-1}^{\beta+1}\right\|^2 = \frac{\pi}{2(\alpha-2m)}, \quad \left\|z_{\alpha-2m-1}^{\beta-1}\right\|^2 = \frac{\epsilon_{\beta-1}\pi}{2(\alpha-2m)}, \quad and \quad \left\|z_{\beta-1}^{\beta-1}\right\|^2 = \frac{\epsilon_{\beta-1}\pi}{2\beta} \quad (28)$$

Where $\epsilon_{\beta-1}=2$ for $\beta=1$ and $\epsilon_{\beta-1}=1$ otherwise. Substituting these into Eq. (27), and simplifying:

$$\begin{cases} if\ \beta > 1:\ \iint\left(\frac{\partial z_\alpha^\beta}{\partial x}\right)^2 \rho d\rho d\theta = \pi\sum_{m=0}^{\frac{\alpha-\beta-2}{2}}(\alpha-2m)+\beta\frac{\pi}{2} = \frac{\pi}{4}(\alpha^2-\beta^2+2\alpha) \\ if\ \beta = 1:\ \iint\left(\frac{\partial z_\alpha^\beta}{\partial x}\right)^2 \rho d\rho d\theta = \frac{3\pi}{2}\sum_{m=0}^{\frac{\alpha-\beta-2}{2}}(\alpha-2m)+\beta\pi = \frac{3\pi}{8}\left(\alpha^2+2\alpha-\frac{1}{3}\right) \end{cases} \quad (29)$$

Using a similar approach, analogous equations for the partial derivatives of Zernike functions can be derived with sine terms with respect to $x$. A comparable process applies to the partial derivatives with respect to $y$ for Zernike functions. However, there are two notable differences between the partial derivatives with respect to $x$ and $y$:

1. As seen in Eqs. (23c-d), the partial derivative of Zernike functions with sine terms with respect to $y$ transforms into Zernike functions with cosine terms, and vice versa. Consequently, the $y$-partial derivative of a Zernike function with subscript $\alpha$ can be expressed as a linear combination of Zernike functions with flipped sine or cosine terms and subscripts smaller than $\alpha$.

2. In transitioning from Eq. (26) to Eq. (27) for the $y$-partial derivative, the cross-multiplication terms are multiplied by $-2$ instead of $+2$. Nevertheless, the integrals of the cross terms vanish in both cases due to the orthogonality of Zernike polynomials. As a result, the same relationships as in Eq. (29) are obtained. The magnitudes of all integral terms will be summarized later in this section of II-C-1.

So far, the cases $\beta=1$ and $\beta>1$ have been addressed. For the final case, $\beta=0$, as discussed earlier, $\beta-1=-1$ is replaced by $|\beta-1|=1$ in Eq. (17), reducing both equations to the following:

$$\left(\frac{d}{d\rho}\right)\left(R_\alpha^0 - R_{\alpha-2}^0\right) = 2\alpha R_{\alpha-1}^1 \quad (30)$$

For $\beta=0$, Zernike polynomials lack an angular component, which causes the $\partial/\partial\theta$ terms in Eq. (16) to vanish. Applying Eq. (30), Eq. (19) simplifies for $\beta=0$ as follows:

$$\frac{\partial z_\alpha^0}{\partial x}-\frac{\partial z_{\alpha-2}^0}{\partial x} = \cos\theta\left[\frac{d}{d\rho}\left(R_\alpha^0 - R_{\alpha-2}^0\right)\right] = 2\alpha\cos\theta\ R_{\alpha-1}^1 = 2\alpha z_{\alpha-1}^1 \quad (31a)$$

$$\frac{\partial z_\alpha^0}{\partial y} - \frac{\partial z_{\alpha-2}^0}{\partial y} = \sin\theta \left[\frac{d}{d\rho}(R_\alpha^0 - R_{\alpha-2}^0)\right] = 2\alpha \sin\theta\, R_{\alpha-1}^1 = 2\alpha z_{\alpha-1}^{-1} \tag{31b}$$

Using the same procedure applied to derive Eq. (25), we can express these partial derivatives as:

$$\frac{\partial z_\alpha^0}{\partial x} = \left[\sum_{m=0}^{\frac{\alpha-2}{2}} 2(\alpha - 2m)(z_{\alpha-2m-1}^1)\right] \tag{32a}$$

$$\frac{\partial z_\alpha^0}{\partial y} = \left[\sum_{m=0}^{\frac{\alpha-2}{2}} 2(\alpha - 2m)(z_{\alpha-2m-1}^{-1})\right] \tag{32b}$$

By applying similar reasoning regarding the orthogonality of Zernike polynomials as utilized in Eq. (27), the following is verified:

$$\iint \left(\frac{\partial z_\alpha^0}{\partial x}\right)^2 \rho\, d\rho\, d\theta = \sum_{m=0}^{\frac{\alpha-\beta-2}{2}} [4(\alpha - 2m)^2 \|z_{\alpha-2m-1}^1\|^2] = \frac{\pi}{2}(\alpha^2 + 2\alpha) \tag{33a}$$

$$\iint \left(\frac{\partial z_\alpha^0}{\partial y}\right)^2 \rho\, d\rho\, d\theta = \sum_{m=0}^{\frac{\alpha-\beta-2}{2}} [4(\alpha - 2m)^2 \|z_{\alpha-2m-1}^{-1}\|^2] = \frac{\pi}{2}(\alpha^2 + 2\alpha) \tag{33b}$$

Next, the normalized versions of the Zernike functions as basis functions are employed, as defined in Eq. (10):

$$\int_0^{2\pi} \int_0^{D/2} \left(\frac{\partial}{\partial x}\left(\frac{Z_\alpha^\mu\left(\frac{2r}{D},\theta\right)}{D\sqrt{\epsilon_\beta \pi(8\alpha+8)}}\right)\right)^2 r\, dr\, d\theta = \frac{\epsilon_\beta \pi(8\alpha+8)}{D^2} \int_0^{2\pi} \int_0^1 \left(\frac{2}{D}\frac{\partial z_\alpha^\mu(\rho,\theta)}{\partial X}\right)^2 \frac{D}{2}\rho\frac{D}{2} d\rho\, d\theta \tag{34}$$

Here, $x = r\cos\theta$, $X = \rho\cos\theta$ and $\beta = |\mu|$. The integral on the right-hand side simplifies to the expressions obtained in Eq. (29). By setting $\beta=|\mu|$ and $\partial u=\partial x$ or $\partial y$, three cases are defined:

$$\int_0^{2\pi} \int_0^{D/2} \left(\frac{\partial}{\partial u}\left(\frac{Z_\alpha^\mu\left(\frac{2r}{D},\theta\right)}{D\sqrt{\epsilon_\beta \pi(8\alpha+8)}}\right)\right)^2 r\, dr\, d\theta = \begin{cases} \frac{2\pi^2(\alpha+1)}{D^2}(\alpha^2 - \beta^2 + 2\alpha), & \text{for } \beta > 1 \\ \frac{3\pi^2(\alpha+1)}{D^2}\left(\alpha^2 + 2\alpha - \frac{1}{3}\right), & \text{for } \beta = 1 \\ \frac{8\pi^2(\alpha+1)}{D^2}(\alpha^2 + 2\alpha), & \text{for } \beta = 0 \end{cases} \tag{35}$$

It is noteworthy that, although this equation and similar ones for other derivatives are not the primary focus of this study, they can be employed to streamline calculations related to RUS when utilizing Zernike polynomials.

### 2. Trace of the matrix $\Gamma$

In this subsection, the formulation for the trace of the matrix $\Gamma$ is derived, corresponding to step (a) of the proof outlined at the end of Section II-C. Expressing the displacement in the Euclidean coordinate system, $\Gamma a$ in Eq.(15) can be written as:

$$\Gamma a = \begin{bmatrix} \Gamma_{11} & \Gamma_{12} & \Gamma_{13} \\ \Gamma_{12} & \Gamma_{22} & \Gamma_{23} \\ \Gamma_{13} & \Gamma_{23} & \Gamma_{33} \end{bmatrix} \begin{bmatrix} a_1 \\ a_2 \\ a_3 \end{bmatrix}, \tag{36}$$

where:

$$\boldsymbol{a_i} = [a_{i1}\ a_{i2}\ a_{i3}\ a_{i4}\ \ldots\ a_{in}]^T, \tag{37}$$

The components $\boldsymbol{\Gamma}_{ij}$ are expressed as follows:

$$\begin{cases} \boldsymbol{\Gamma_{11}} = C_{11}\boldsymbol{\Phi_{11}} + C_{61}\boldsymbol{\Phi_{21}} + C_{51}\boldsymbol{\Phi_{31}} + C_{16}\boldsymbol{\Phi_{12}} + C_{66}\boldsymbol{\Phi_{22}} + C_{56}\boldsymbol{\Phi_{32}} + C_{15}\boldsymbol{\Phi_{13}} + C_{65}\boldsymbol{\Phi_{23}} + C_{55}\boldsymbol{\Phi_{33}} \\ \boldsymbol{\Gamma_{22}} = C_{66}\boldsymbol{\Phi_{11}} + C_{26}\boldsymbol{\Phi_{21}} + C_{46}\boldsymbol{\Phi_{31}} + C_{62}\boldsymbol{\Phi_{12}} + C_{22}\boldsymbol{\Phi_{22}} + C_{42}\boldsymbol{\Phi_{32}} + C_{64}\boldsymbol{\Phi_{13}} + C_{24}\boldsymbol{\Phi_{23}} + C_{44}\boldsymbol{\Phi_{33}} \\ \boldsymbol{\Gamma_{33}} = C_{55}\boldsymbol{\Phi_{11}} + C_{45}\boldsymbol{\Phi_{21}} + C_{35}\boldsymbol{\Phi_{31}} + C_{54}\boldsymbol{\Phi_{12}} + C_{44}\boldsymbol{\Phi_{22}} + C_{34}\boldsymbol{\Phi_{32}} + C_{53}\boldsymbol{\Phi_{13}} + C_{43}\boldsymbol{\Phi_{23}} + C_{33}\boldsymbol{\Phi_{33}} \end{cases} \tag{38}$$

In Eq. (38), $C_{ij}$ represent the elastic constants in Voigt notation, and the matrices $\Phi_{ij}$ are defined as:

$$\boldsymbol{\Phi_{ij}} = \begin{bmatrix} \int_V \varphi_{1,i}\,\varphi_{1,j}\,dV & \int_V \varphi_{1,i}\,\varphi_{2,j}\,dV & \int_V \varphi_{1,i}\,\varphi_{3,j}\,dV & \cdots & \int_V \varphi_{1,i}\,\varphi_{s,j}\,dV & \cdots \\ \int_V \varphi_{2,i}\,\varphi_{1,j}\,dV & \int_V \varphi_{2,i}\,\varphi_{2,j}\,dV & \int_V \varphi_{2,i}\,\varphi_{3,j}\,dV & \cdots & \int_V \varphi_{2,i}\,\varphi_{s,j}\,dV & \cdots \\ \vdots & \vdots & \vdots & \vdots & \vdots & \vdots \\ \int_V \varphi_{r,i}\,\varphi_{1,j}\,dV & \int_V \varphi_{r,i}\,\varphi_{2,j}\,dV & \int_V \varphi_{r,i}\,\varphi_{3,j}\,dV & \cdots & \int_V \varphi_{r,i}\,\varphi_{s,j}\,dV & \cdots \\ \vdots & \vdots & \vdots & \vdots & \vdots & \vdots \end{bmatrix} \tag{39}$$

Here, $\varphi_r$'s are the basis functions defined in Eq. (10), and the subscript ,i denotes the partial derivative, i.e., $\varphi_{r,i} = \frac{\partial \varphi_r}{\partial x_i}$.

The intermediate goal is to determine the sum of the eigenvalues of $\boldsymbol{\Gamma}$ in Eq. (15), which is equivalent to the trace of $\boldsymbol{\Gamma}$. From Eq. (38), calculating the trace of $\boldsymbol{\Gamma}$ requires evaluating the trace of the $\boldsymbol{\Phi}_{ij}$ matrices defined in Eq. (39).

### 3. Analyzing the trace of the matrix $\Gamma$

In this subsection, steps (b) and (c) of the proof outlined at the end of Section II-C are addressed. Specifically, an analysis of how the trace of $\boldsymbol{\Gamma}$ changes when transitioning from N-1 to N is presented.

The trace of $\boldsymbol{\Gamma}$ is determined by the traces of $\boldsymbol{\Gamma_{11}}$, $\boldsymbol{\Gamma_{22}}$ and $\boldsymbol{\Gamma_{33}}$. According to Eq. (38), these traces depend on the elastic constants and the traces of the $\boldsymbol{\Phi}_{ij}$ matrices. The traces of the $\boldsymbol{\Phi}_{ij}$ matrices consist of integrals of the form:

$$\int_V \varphi_{q,i}\,\varphi_{q,j}\,dV \tag{40}$$

First, the integral in Eq. (40) for the case $i \neq j$ is considered. To begin, consider the scenario where $i$ and $j$ correspond to 1 and 2, i.e., within the plane of the base of the cylinder. Without loss of generality, $i=1$ and $j=2$ is assumed. Under this assumption, Eq. (40) can be rewritten as:

$$\int_V \left( \frac{\frac{\partial z_\alpha^\mu}{\partial x}\left(\frac{2r}{D},\theta\right)}{\sqrt{\epsilon_\beta \pi (8\alpha + 8)}} \right) \left( \frac{P_\gamma\left(\frac{2x_3}{D_3}\right)}{\sqrt{\frac{D_3}{2\gamma+1}}} \right) \left( \frac{\frac{\partial z_\alpha^\mu}{\partial y}\left(\frac{2r}{D},\theta\right)}{\sqrt{\epsilon_\beta \pi (8\alpha + 8)}} \right) \left( \frac{P_\gamma\left(\frac{2x_3}{D_3}\right)}{\sqrt{\frac{D_3}{2\gamma+1}}} \right) dV \tag{41}$$

Focusing on the first and third terms of the integral, as discussed in Section II-C-1, both the $x$ and $y$ partial derivatives of the Zernike function with subscript $\alpha$ can be expressed as linear combinations of Zernike functions with subscripts smaller than $\alpha$. Furthermore, if the Zernike polynomial involves a sine function (i.e., $\mu > 0$), the $x$-partial derivative is a sum of Zernike polynomials with sine functions, while the $y$-partial derivative is a sum of Zernike polynomials with cosine functions. Consequently, two sums of functions where each element in one sum is orthogonal to all elements in the other are obtained. As a result, the integral vanishes.

Considering the case where $i=1$ and $j=3$, Eq. (40) can be rewritten as:

$$\int_V \left( \frac{\frac{\partial z_\alpha^\mu}{\partial x}\left(\frac{2r}{D},\theta\right)}{\sqrt{\epsilon_\beta \pi (8\alpha+8)}} \right) \left( \frac{P_\gamma\left(\frac{2x_3}{D_3}\right)}{\sqrt{\frac{D_3}{2\gamma+1}}} \right) \left( \frac{Z_\alpha^\mu\left(\frac{2r}{D},\theta\right)}{\sqrt{\epsilon_\beta \pi (8\alpha+8)}} \right) \left( \frac{\frac{d}{d_3}P_\gamma\left(\frac{2x_3}{D_3}\right)}{\sqrt{\frac{D_3}{2\gamma+1}}} \right) dV \quad (42)$$

In this case, the final term, which represents the derivative of the Legendre polynomial, is one degree lower than the polynomial itself. Consequently, it is orthogonal to the polynomial, resulting in a zero volume integral. The same reasoning applies when $i=2$ and $j=3$.

The only scenario where the integral does not vanish is when $i=j$, i.e., when the derivative is multiplied by itself within the integral. To analyze the contributions to the trace when increasing the number of basis functions from N-1 to N, the case where $\alpha+\gamma=N$ is considered. The additional elements contributing to the trace of $\mathbf{\Phi}_{ii}$ can be expressed as:

$$\sum_{q=\frac{(N)(N+1)(N+2)}{6}}^{\frac{(N+1)(N+2)(N+3)}{6}} \int_V \varphi_{q,i} \, \varphi_{q,i} \, dV. \quad (43)$$

For $i=3$, the integral is given by Eq. (39) and Eq. (42) in our previous work [39]:

$$\frac{1}{D_3^2}\left[\frac{N^5}{5} + \frac{3N^4}{2} + \frac{4N^3}{1} + \frac{9N^2}{2} + \frac{9N}{5}\right]. \quad (44)$$

Next, Eq. (35) can be applied to calculate the derivatives for $i=1, 2$. As previously explained, the integral term in Eq. (42) is identical for $x$ and $y$-partial derivatives, so it suffices to compute the integral for $i=1$. The number of basis functions $\boldsymbol{\varphi}_q$ corresponding to $\beta=0$, $\beta=1$ and $\beta>1$ is determined next.

**Case 1:** $\beta=0$

When $\beta=0$, $\alpha$ ranges from 0 to N but only includes even numbers. For each $\alpha$, there is a single corresponding value of $\gamma$, namely $N-\gamma$. The sum becomes:

$$\sum_{\alpha=0,2,4,\dots}^{N} \frac{8\pi^2(\alpha+1)}{D^2}(\alpha^2+2\alpha) = \begin{cases} N \text{ even: } \sum_{m=0,1,2,\dots}^{N/2} \frac{32\pi^2(2m+1)}{D^2}(m^2+m) = \frac{\pi^2}{D^2}(N^4+8N^3+20N^2+16N) \\ N \text{ odd: } \sum_{m=0,1,2,3,\dots}^{(N-1)/2} \frac{32\pi^2(2m+1)}{D^2}(m^2+m) = \frac{\pi^2}{D^2}\left(\frac{1}{16}N^4+\frac{3}{4}N^3+\frac{19}{8}N^2+\frac{3}{4}N+\frac{63}{16}\right) \end{cases} \quad (45)$$

In which the sum depends on whether N is even or odd.

**Case 2:** $\beta=1$

For $\beta=1$, there are two values of $\mu$: $\mu=+1$ (sine terms) and $\mu=-1$ (cosine terms). Here, $\alpha$ ranges from 1 to N, including only odd numbers. For each $\alpha$, there is a unique value of $\gamma$, leading to:

$$2\sum_{\alpha=1,3,\dots}^{N} \frac{3\pi^2(\alpha+1)}{D^2}\left(\alpha^2+2\alpha-\frac{1}{3}\right) = \begin{cases} N \text{ even: } \sum_{m=0,1,2,\dots}^{N/2} \frac{24\pi^2(2m+1)}{D^2}\left(m^2+m-\frac{1}{12}\right) = \frac{\pi^2}{D^2}\left(\frac{3}{4}N^4+6N^3+\frac{29}{2}N^2+10N\right) \\ N \text{ odd: } \sum_{m=0,1,2,3,\dots}^{(N-1)/2} \frac{24\pi^2(2m+1)}{D^2}\left(m^2+m-\frac{1}{12}\right) = \frac{\pi^2}{D^2}\left(\frac{1}{16}N^4+\frac{3}{4}N^3+\frac{53}{24}N^2-\frac{11}{12}N-\frac{101}{48}\right) \end{cases} \quad (46)$$

The factor of 2 accounts for the two values of $\mu$ for $\beta=1$.

**Case 3:** $\beta>1$

For $\beta>1$, the relationship between $\alpha$ and $\beta$ ($\alpha-\beta$ must be even) introduces additional complexity. The sum depends on whether N and $\beta$ are even or odd, requiring a division into subcases. Each subcase can be addressed separately based on these parity conditions:

$$N \text{ even} \begin{cases} \beta = 2k, \alpha = 2l: \sum_{l=1}^{N/2} \sum_{k=1}^{l} 2 \times \frac{2\pi^2(2l+1)}{D^2}(4l^2 - 4k^2 + 4l) = \frac{\pi^2}{D^2}\left(\frac{2}{15}N^5 + \frac{13}{12}N^4 + \frac{8}{3}N^3 + \frac{5}{3}N^2 - \frac{4}{5}N\right) \\ \beta = 2k+1, \alpha = 2l+1: \sum_{l=1}^{\frac{N-2}{2}} \sum_{k=1}^{l} 2 \times \frac{2\pi^2(2l+1+1)}{D^2}(4l^2 - 4k^2 - 4k + 8l) = \frac{\pi^2}{D^2}\left(\frac{2}{15}N^5 + \frac{1}{6}N^4 - 2N^3 - \frac{2}{3}N^2 + \frac{88}{15}N\right) \end{cases}$$
(47a)

$$N \text{ odd} \begin{cases} \beta = 2k, \alpha = 2l: \sum_{l=1}^{\frac{N-1}{2}} \sum_{k=1}^{l} 2 \times \frac{2\pi^2(2l+1)}{D^2}(4l^2 - 4k^2 + 4l) = \frac{\pi^2}{D^2}\left(\frac{2}{15}N^5 + \frac{5}{12}N^4 - \frac{1}{3}N^3 - \frac{7}{6}N^2 + \frac{1}{5}N + \frac{3}{4}\right) \\ \beta = 2k+1, \alpha = 2l+1: \sum_{l=1}^{\frac{N-1}{2}} \sum_{k=1}^{l} 2 \times \frac{2\pi^2(2l+1+1)}{D^2}(4l^2 - 4k^2 - 4k + 8l) = \frac{\pi^2}{D^2}\left(\frac{2}{15}N^5 + \frac{5}{6}N^4 - \frac{13}{3}N^2 - \frac{2}{15}N + \frac{7}{2}\right) \end{cases}$$
(47b)

By summing over all cases of β for both even and odd N, as described in Eqs. (45–47), the sum in Eq. (43) for i=1 or 2 is obtained:

$$\begin{cases} N \text{ even}: \sum_{q=\frac{(N)(N+1)(N+2)}{6}}^{\frac{(N+1)(N+2)(N+3)}{6}} \int_V \varphi_{q,i} \, \varphi_{q,i} \, dV = \frac{\pi^2}{D^2}\left(\frac{4}{15}N^5 + 3N^4 + \frac{44}{3}N^3 + \frac{71}{2}N^2 + \frac{466}{15}N\right) \\ N \text{ odd}: \sum_{q=\frac{(N)(N+1)(N+2)}{6}}^{\frac{(N+1)(N+2)(N+3)}{6}} \int_V \varphi_{q,i} \, \varphi_{q,i} \, dV = \frac{\pi^2}{D^2}\left(\frac{4}{15}N^5 + \frac{11}{8}N^4 + \frac{7}{6}N^3 - \frac{11}{12}N^2 - \frac{1}{10}N - \frac{43}{24}\right) \end{cases}$$
(48)

It can be readily observed that the right-hand side of Eq. (48) is larger when N is even.

Referring to Eq. (38) for $\Gamma_{11}$, $\Gamma_{22}$ and $\Gamma_{33}$, each term contains contributions involving elastic constants and $\Phi_{ii}$. When N−1 is incremented to N, Eq. (44) provides the added contribution to $\Phi_{33}$, while Eq. (48) describes the additional contributions to $\Phi_{11}$ and $\Phi_{22}$. Thus, the total added value to the trace of $\Gamma$ when N−1 is incremented to N can be expressed as:

$$\begin{cases} N \text{ even}: b_5 N^5 + b_{4e} N^4 + b_{3e} N^3 + b_{2e} N^2 + b_{1e} N + b_{0e} \\ N \text{ odd}: b_5 N^5 + b_{4o} N^4 + b_{3o} N^3 + b_{2o} N^2 + b_{1o} N + b_{0o} \end{cases}$$
(49)

Here, the coefficients $b_5$, $b_{io}$, $b_{ie}$'s are determined by the diagonal elements of the elastic constants in Eq. (38) and the numerical factors derived from Eqs. (44) and (48). Notably, $b_5$ is the same for both even and odd N.

The reasoning outlined in Section II-C can now be applied. When the number of basis functions is increased from N−1 to N, the size of the matrix $\Gamma$ increases by:

$$\frac{3(N+1)(N+2)(N+3)}{6} - \frac{3(N)(N+1)(N+2)}{6} = \frac{3N^2 + 9N + 6}{2}$$
(50)

The difference between the traces of the larger and smaller matrices $\Gamma$ is given by Eq. (49). This difference is also equal to the sum of the added eigenvalues. Since the largest eigenvalue is $\lambda_n$, and given that the sum for even N is greater than for odd N (as can be verified by Eq. (49)), a trace inequality is established:

$$b_5 N^5 + b_{4o} N^4 + b_{3o} N^3 + b_{2o} N^2 + b_{1o} N + b_{0o} \le \left(\frac{3N^2 + 9N + 6}{2}\right) \lambda_n$$
(51)

If the number of basis functions are further increased, from N to N+1, the added dimension of $\Gamma$ and the added contribution to the trace of $\Gamma$ can be determined by substituting N with N+1 in Eqs. (49) and (50):

$$\begin{cases} N \text{ even}: b_5 N^5 + h_{4e} N^4 + h_{3e} N^3 + h_{2e} N^2 + h_{1e} N + h_{0e} \\ N \text{ odd}: b_5 N^5 + h_{4o} N^4 + h_{3o} N^3 + h_{2o} N^2 + h_{1o} N + h_{0o} \end{cases}$$
(52a)

$$\frac{3(N+2)(N+3)(N+4)}{6} - \frac{3(N+1)(N+2)(N+3)}{6} = \frac{3N^2+15N+18}{2} \tag{52b}$$

As before, the coefficients $b_5$, $h_{io}$, $h_{ie}$'s are derived from the diagonal elements of the elastic constants and the numerical factors from Eqs. (44) and (48). Importantly, while $b_5$ remains constant, the other coefficients vary. Additionally, the leading coefficient governing the change in the dimension of $\Gamma$ also remains constant. Since all new eigenvalues arising from the increased dimension of $\Gamma$ are greater than $\lambda_n$, the following inequality is defined:

$$\left(\frac{3N^2+15N+18}{2}\right)\lambda_n \leq b_5 N^5 + h_{4e} N^4 + h_{3e} N^3 + h_{2e} N^2 + h_{1e} N + h_{0e} \tag{53}$$

Combining Eq. (53) with Eq. (51) and dividing by n,:

$$\frac{b_5 N^5 + b_{4o} N^4 + b_{3o} N^3 + b_{2o} N^2 + b_{1o} N + b_{0o}}{\left(\frac{3N^2+9N+6}{2}\right)\left(\frac{3(N+3)(N+2)(N+1)}{6}\right)} \leq \frac{\lambda_n}{n} \leq \frac{b_5 N^5 + h_{4e} N^4 + h_{3e} N^3 + h_{2e} N^2 + h_{1e} N + h_{0e}}{\left(\frac{3N^2+15N+18}{2}\right)\left(\frac{3(N+3)(N+2)(N+1)}{6}\right)} \tag{54}$$

Which simplifies further as:

$$\frac{b_5 N^5 + b_{4o} N^4 + b_{3o} N^3 + b_{2o} N^2 + b_{1o} N + b_{0o}}{0.75 N^5 + 6.75 N^4 + 23.25 N^3 + 38.25 N^2 + 30 N + 9} \leq \frac{\lambda_n}{n} \leq \frac{b_5 N^5 + h_{4e} N^4 + h_{3e} N^3 + h_{2e} N^2 + h_{1e} N + h_{0e}}{0.75 N^5 + 8.25 N^4 + 35.25 N^3 + 72.75 N^2 + 72 N + 27} \tag{55}$$

Taking the limit of both the left-hand and right-hand sides as N (or n) approaches infinity, both converge to the same value, which is:

$$\lim_{n \to \infty} \frac{\lambda_n}{n} = \lim_{n \to \infty} \frac{\rho \omega_n^2}{n} = \frac{4 b_5}{3} \tag{56}$$

This result indicates that the constant depends on the diagonal elements of $\Gamma$. As demonstrated in Eqs. (22) and subsection II-B-4, this constant ultimately relies on the diagonal elements of the elastic constant matrix.

### III.  NUMERICAL RESULTS AND DISCUSSIONS

For the numerical calculations, verification that the asymptotic behavior, as described in Eqs. (44) and (48), is inversely proportional to the square of the sample's diameter and height is the goal. Additionally, as discussed previously, the trace of $\Gamma$ in Eq. (38) is influenced solely by the diagonal elements of the elastic constants. To test this, two samples are selected: one with dimensions twice that of the other and elastic constants adjusted accordingly (four times as much), such that the asymptotes for both samples should converge.

The first sample has a diameter of 5 mm, a height of 7.5 mm, and elastic constants of anisotropic iron [46] with $C_{11}$=228 GPa, $C_{12}$=133 GPa, and $C_{44}$=111 GPa. The second sample, with a diameter of 10 mm and a height of 15 mm (twice the dimensions of the first), has elastic constants $C_{11}$=912 GPa and $C_{44}$=444 GPa, while $C_{12}$ remains the same at 133 GPa.

In this analysis, $\lambda_n = \rho \omega^2$, where $\omega$ represents the angular frequency. Plots of $\lambda_n/n$ for both samples across various values of N are presented in Fig 3. As noted in [39], the eigenvalues calculated using a finite basis are greater than the actual resonant frequencies. This phenomenon is evident in Fig. 3, where the slope of $\lambda n/n$ diminishes more rapidly as N increases and stabilizes over a wider range.

In Section II-C-3, the mathematical convergence of eigenvalues for large values of n was demonstrated. While computational limitations prevented simulations for very large n, the asymptotes for both cases converge for all values of N were observed, albeit at different rates.

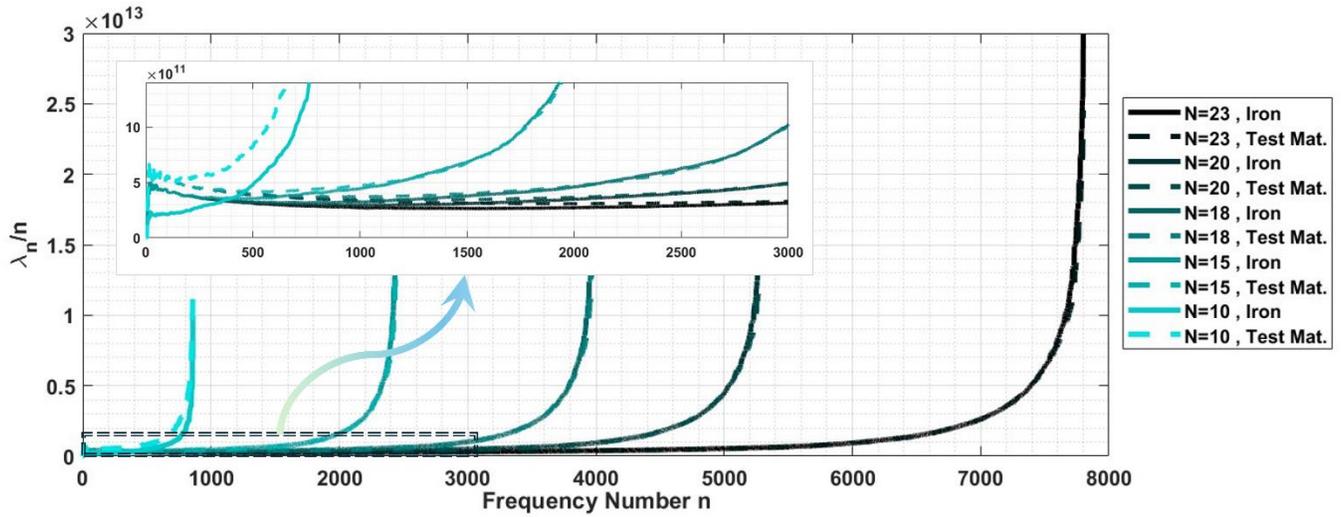

Figure 3: Plot of $\lambda n/n$ for all frequency numbers for two samples across different values of N. The dimensions and elastic constants of the two samples are chosen such that, based on our analysis, their asymptotic values should converge.

## IV. CONCLUSIONS

This study explored the asymptotic behavior of eigenfrequencies in cylindrical samples using RUS. Building on prior research, a framework that leverages Zernike and Legendre polynomials was developed to analyze resonant frequencies in samples with cylindrical geometry. Findings revealed that the trace of the matrix $\Gamma$ and its eigenvalues exhibit predictable growth patterns as the basis set is expanded, with specific dependencies on the diagonal elastic constants of the sample. Importantly, the asymptotic convergence of eigenvalue ratios highlights that greater frequencies provide diminishing returns in terms of new information, aligning with observations in cuboid geometries in previous work. This work enhances the theoretical understanding of RUS while providing practical insights into optimizing its application for cylindrical samples. Future research may extend these results to samples of arbitrary geometry, further generalizing the methodology and its implications.